\documentclass[letterpaper, 10 pt, conference]{ieeeconf}

\usepackage[utf8]{inputenc}
\usepackage{amsmath,bm} 
\usepackage{amssymb}
\usepackage{graphicx}
\usepackage{blindtext}
\usepackage{wrapfig}
\usepackage{subcaption}
\usepackage{color,soul}
\usepackage{cite}

\begin{document}

\title{Affine Transformation-based Perfectly Undetectable False Data Injection Attacks on Remote Manipulator Kinematic Control with Attack Detector}


\author
{\authorblockN{Jun Ueda\authorrefmark{1} and  Jacob Blevins\authorrefmark{1}}
\authorblockA{\authorrefmark{1}The George W. Woodruff School of Mechanical Engineering, Georgia Institute of Technology,\\
771 Ferst Drive, Atlanta, GA 30332-0405, USA. E-mail: jun.ueda@me.gatech.edu, jacob.blevins@gatech.edu\\}}


\markboth{IEEE Robotics and Automation Letters. Preprint Version. Accepted XXX}
{Ueda \MakeLowercase{\textit{et al.}}: fully stealthy false data injection attacks} 

\maketitle

\begin{abstract}
This paper demonstrates the viability of perfectly undetectable affine transformation attacks against robotic manipulators where intelligent attackers can inject multiplicative and additive false data while remaining completely hidden from system users. The attacker can implement these communication line attacks by satisfying three Conditions presented in this work. These claims are experimentally validated on a FANUC 6 degree of freedom manipulator by comparing a nominal (non-attacked) trial and a detectable attack case against three perfectly undetectable trajectory attack Scenarios: scaling, reflection, and shearing. The results show similar observed end effector error for the attack Scenarios and the nominal case, indicating that the perfectly undetectable affine transformation attack method keeps the attacker perfectly hidden while enabling them to attack manipulator trajectories.
\end{abstract}

{\fontsize{9}{12}\selectfont
\textbf{\textit{Index Terms} -- False data injection attack, Jacobian velocity control, Affine transformation}
}


\section{Introduction}

\begin{figure*}[ht]
	\centering
	\includegraphics[width=1.5\columnwidth]{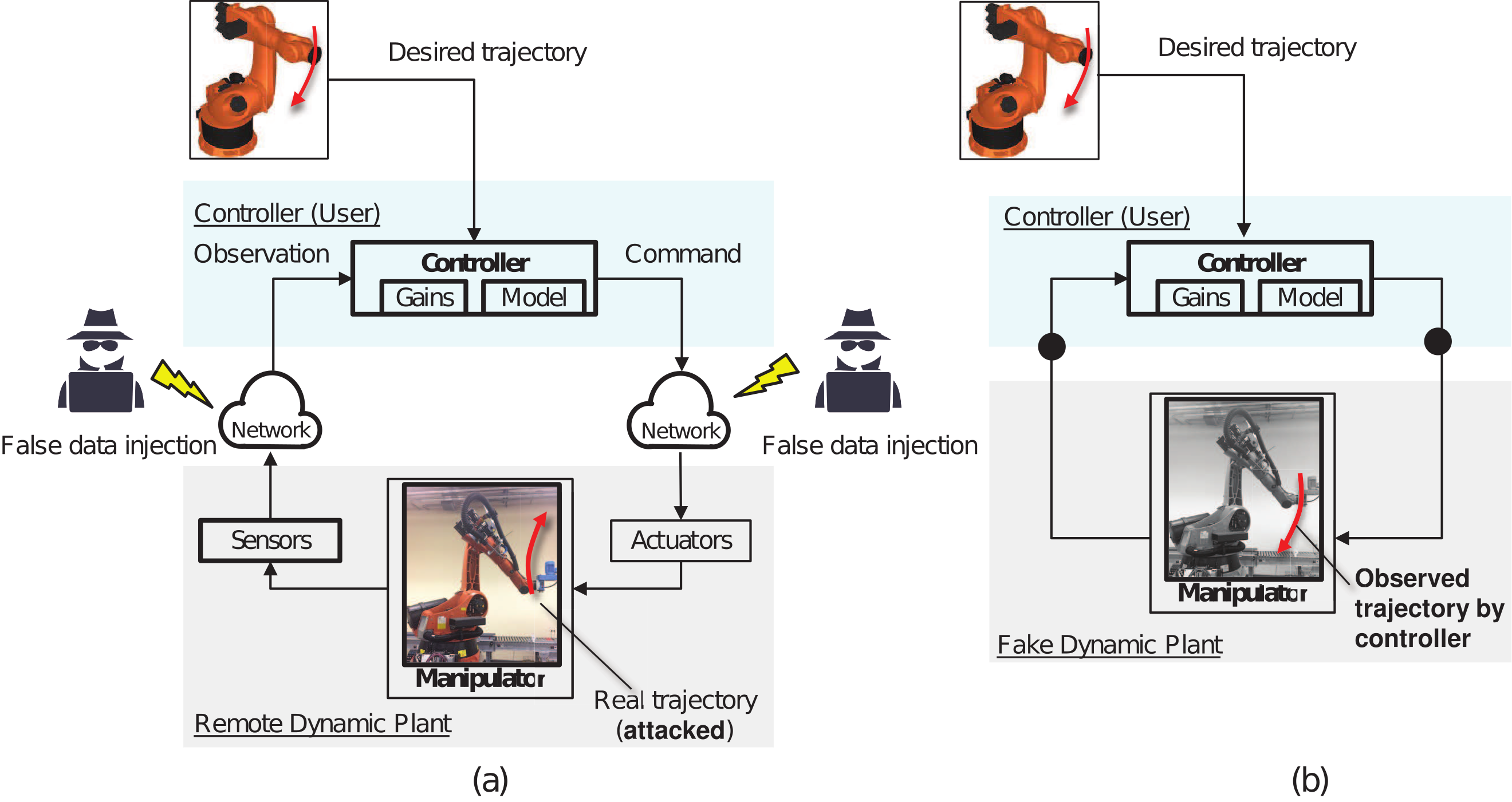}
	\caption{Conceptual diagram of false data injection attack (FDIA) on remote manipulator: (a) Attacked networked control system with coordinated FDIA on the commands and observables. (b) Plant dynamics as perceived by the controller, indistinguishable from the nominal plant behavior and thus undetectable.}
	\label{FDIA_manipulator_concept}
\end{figure*}

In the era of Industry 4.0, virtually all modern devices are interconnected via the internet, facilitating the exchange of sensor measurements, control commands, and other vital information for monitoring and controlling complex systems through computer networks \cite{cyber-physical}. The cybersecurity of these communication channels is increasingly under threat from various adversaries, raising significant concerns\cite{survey}. A typical networked control system is illustrated in Fig. \ref{FDIA_manipulator_concept}, wherein a physical plant such as an industrial robotic manipulator is connected to the internet, allowing it to be controlled remotely by the user. The controller receives measurements from the plant via a communication channel and computes control commands that are sent back to the plant via another communication channel, forming a closed-loop system for remote control of the manipulator. 

One notable form of cybersecurity threat is the False Data Injection Attack (FDIA), wherein adversaries tamper with sensor, control, or monitoring signals within the communication lines to adversely affect the behavior, performance, and stability of the targeted system \cite{Sandberg22}. Rather than gaining unauthorized access to the controller or the plant (including its local, low-level embedded controllers), conducting FDIA on communication lines is likely a more efficient and secretive approach for adversaries looking to disrupt operations. 

There are three types of FDIAs: detectable, undetectable, and stealthy \cite{Sandberg22}. Detectable FDIAs are attacks in which the controller can observe an active injection attack; this form of attack is well studied and can be easily rejected or compensated for. Current literature for detectable attacks suggests isolating observed anomalies to determine if there is an attack and to understand its type and the affected channels. If the attack is detected, a fault can be thrown \cite{Sandberg22}. Other methods for defending against detectable attacks include comparing sensor values against pre-determined models or expected values and adapting the control input as if the attack is a disturbance \cite{TERANISHI2019297}. Additionally, adaptive control can be used to estimate a scalar multiplicative attack constant and reject it in the control law \cite{Zhang23}.

Undetectable and stealthy attacks are characterized by attacks which are more difficult for the operator to detect. In the case of an undetectable attack, the attacked signals coincide with those that are within regular operating range, thus faults and regular detectors fail \cite{Pasqualetti13}. Perfectly undetectable attacks are those where there is no change in observed states, but the system is still attacked. A relative of the undetectable attack is the stealthy attack. Stealthy attacks are those where an attack detector is explicitly taken into account; therefore, ensuring that the attacker does not set off an alarm \cite{Sandberg22}.



The significance of this research is the formulation of a generalized FDIA that involves coordinated multiplicative and additive data injections on both control commands and observables, allowing attackers to create perfectly undetectable attacks against kinematic manipulator systems. As far as the authors are aware, current literature studies either additive or multiplicative FDIA on control commands or observables, and often not for perfectly undetectable attacks\cite{Zhu23, LIU2016708}. Covert attack literature has been aimed at both commands and observables and may include multiplication and addition, but just for stealthy FDIAs \cite{Schellenberger}. By taking remote manipulator kinematic control as a representative example, which is also considered in FDIA literature \cite{Zhang23}, this paper demonstrates that the specific structure of the plant dynamics, from commands to observables, allows for a range of perfectly undetectable FDIAs, {\it regardless} of the type of feedback control and attack detector.

\section{Perfectly undetectable FDIA from the controller's perspective}

\subsection{FDIA on networked control systems}
In game theory, an attacker's strategy for a perfectly undetectable attack is centered on maximizing the impact of the attack while minimizing the risk of detection \cite{Manshaei13}. The attacker selects target systems and attack methods that can bypass the defender's measures, weighing the impact of the attack against the risks associated with detection \cite{Roy10}.  This approach requires a deep understanding of the defender's counteractions and the awareness of the security landscape \cite{Zhu13}. One of the key aspects of this attack is the utility function, which quantifies the attacker's strategy, and risk tolerances, guiding their decisions towards optimizing the effectiveness of the attack with minimal exposure \cite{pawlick2019gametheoretic}.

When FDIA on a networked control system is concerned, as shown in Fig. \ref{FDIA_manipulator_concept}(a), common attack strategies include sensor spoofing, destabilization, and performance degradation, each aiming to elicit responses from the plant that deviate from those observed under normal operation. 
While FDIA on control commands is chosen based on specific objectives, an intelligent attacker would avoid using solely command attacks that lead to performance degradation \cite{Zhang23} or immediate instability \cite{TERANISHI2019297}. Such command attacks, due to their noticeable impact on observables, are easily detectable.

Instead, an intelligent attacker would opt for an attack that is difficult to detect or remains perfectly undetected by the users \cite{Sandberg22,Milošević20, Schellenberger, Mao20}. 
Perfectly undetectable attacks from the plant's perspective are defined in the literature, as shown in Appendix \ref{appen_perfectFDIAplant} for interested readers' reference. It must be mentioned that, in this definition, it is assumed that the detector receives ground truth observables without being compromised by FDIA. 

\subsection{Perfectly undetectable FDIA via coordinated attacks on commands and observables from the controller's perspective}

\begin{figure}[th]
	\centering
	\includegraphics[width=.85\columnwidth]{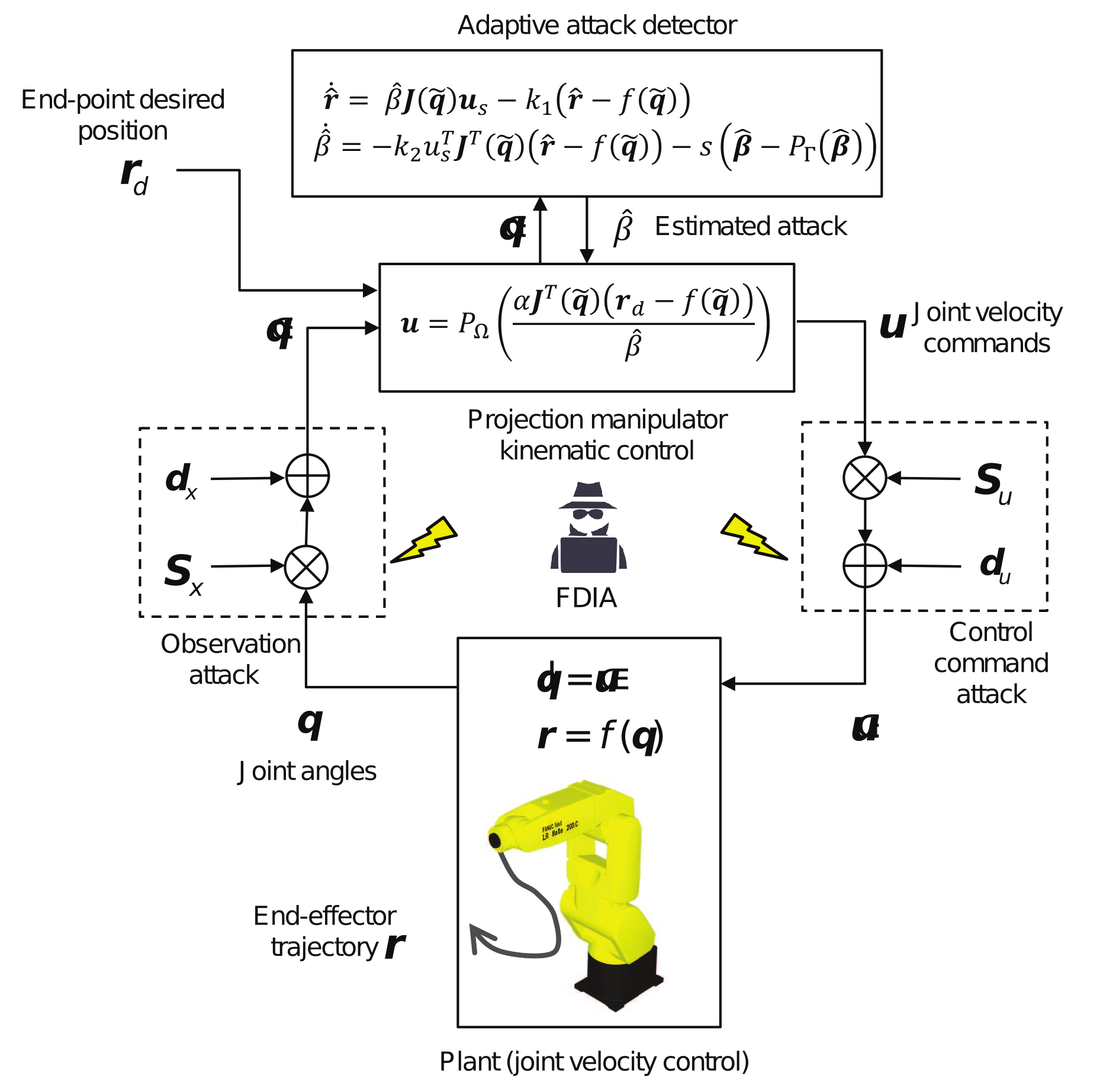}
	\caption{FDIA on manipulator kinematic control: Attack detector and control scheme adopted from \cite{Zhang23}. }
	\label{FDIA_manipulator}
\end{figure}

This paper adopts methods in \cite{Zhang23} for FDIA on networked manipulator kinematic control, and we define perfectly undetectable attacks from the perspective of the controller, detailing specifications related to the networked control architecture illustrated in Fig. \ref{FDIA_manipulator_concept}.
Illustrated in Fig. \ref{FDIA_manipulator}, typical Jacobian-based joint velocity control is  where $\bm{q}=[\theta_1, \theta_2,\cdots,]^T \in R^n$ is a joint angle vector, 
$\bm{r}$ is the end-point position and orientation in the 3D space, typically, $\bm{r}\in R^6$, $\bm{r}=f(\bm{q})$ is forward kinematics, and 
$\bm{J}(\bm{q}) = \frac{\partial f}{\partial \bm{q}^T}$ is the Jacobian matrix.  
With the desired end-effector trajectory $\bm{r}_d$, commonly, a simple kinematic control law utilizing Jacobian transpose is given as:
\begin{equation}
    \bm{u}= \bm{J}(\bm{q})^T (\bm{r}_d - \bm{r}),
\end{equation}
\noindent 
where $\bm{u} \in R^m$ is a control velocity command. Variations of control implementations will be discussed in Section \ref{discussion}. 

A generalized form of FDIA that involves coordinated multiplicative and additive data injections into both control commands and observables is represented in Fig. \ref{FDIA_manipulator} by introducing affine transformations. 
$\tilde{\bm{q}}$ is a compromised observables vector resulting from the attack, $\bm{S}_x \in R^{n \times n}$ represents an arbitrary transformation, such as scaling, reflection, and shearing, and  $\bm{d}_x \in R^n$ represents a translation introducing an offset. This paper assumes $\bm{S}_x$ and $\bm{d}_x$ are constants. 
The affine transformation attack to the observables is

\begin{equation}
    {\bm{\tilde q}} = \bm{\alpha}(\bm{q}) =\bm{S}_x \bm{q} + {{\bm{d}}_x}, \label{tildeq}
\end{equation}
\noindent where $\bm{\alpha}$ is a static observables attack function, and alternatively, in the form of homogeneous transformation:
\begin{equation}
\left[ {\begin{array}{*{20}{c}}
{{\bm{\tilde q}}}\\
1
\end{array}} \right] = \left[ {\begin{array}{*{20}{c}}
{{{\bm{S}}_x}}&{{{\bm{d}}_x}}\\
0&1
\end{array}} \right]\left[ {\begin{array}{*{20}{c}}
{\bm{q}}\\
1
\end{array}} \right]
\end{equation}
\noindent

Similarly, to the control command, $\tilde{\bm{u}}$ is a compromised (attacked) control command vector resulting from the attack by a static attack function, $\bm{\beta}$. When $\bm{\beta}$ is linear affine transformation, $\bm{S}_u \in R^{m \times m}$ represents an arbitrary transformation and  $\bm{d}_u \in R^m$ represents a translation introducing an offset. This paper assumes that $\bm{S}_u$ and $\bm{d}_u$ are constants.

\begin{equation}
{\bm{\tilde u}} = \bm{\beta}(\bm{u})={{\bm{S}}_u}{\bm{u}} + {{\bm{d}}_u}
\label{tildeu}
\end{equation}

\begin{equation}
\left[ {\begin{array}{*{20}{c}}
{{\bm{\tilde u}}}\\
1
\end{array}} \right] = \left[ {\begin{array}{*{20}{c}}
{{{\bm{S}}_u}}&{{{\bm{d}}_u}}\\
0&1
\end{array}} \right]\left[ {\begin{array}{*{20}{c}}
{\bm{u}}\\
1
\end{array}} \right]
\end{equation}

Most studies have considered either additive or multiplicative FDIA on control commands or observables.
For example, 
in \cite{Zhang23}, a multiplicative attack on the control command with a scalar and constant attack parameter $\beta$ was considered, i.e., $\bm{S}_u=\beta \bm{I}^{m \times m}, \bm{d}_u=0, \bm{S}_x= \bm{I}^{n \times n}, \bm{d}_x=0$.
Similarly, \cite{Zhu23} studied both multiplicative and additive data injections, but the observables remained uncompromised.
Representing FDIAs in the form of affine transformations with (\ref{tildeq}) and  (\ref{tildeu}) allows for more generalized analyses. 
While this paper assumes constants for $\bm{S}_u, \bm{d}_u, \bm{S}_x, \bm{d}_x$, 
time-variant attacks will be considered in future publications. In comparison to time varying covert attacks, the linear affine transformation FDIA is simple for attackers to implement since it is a static attack, giving this form of attack greater significance.

This paper considers FDIA being undetectable when the user on the controller side is not able to distinguish the difference between the observed and predicted output estimated based on the nominal plant dynamics with the same control policy. 
The actual plant trajectory is altered by a control command attack. However, when the controller still observes the unaltered trajectory due to an observation attack, an attack detector in the controller may fail to detect the attack. An intelligent attacker coordinates both the control command attack and observation attack to make the attack perfectly undetectable. This definition is not in contradiction with (\ref{perfectFDIAplantdef}) or those given in the literature \cite{Sandberg22,Milošević20,Mao20}.

{\bf Definition 1 (Perfectly undetectable FDIA from the controller's perspective)}: Let $\bm{x}(t,\bm{x}(0),u,\bm{\alpha},\bm{\beta})$ denotes the state variables of a dynamic plant with the initial condition $\bm{x}(0)$, state-feedback controller $\bm{u}=k(\bm{x})$, and FDIA attack functions $\bm{\alpha}(\bm{x})$,$\bm{\beta}(\bm{u})$. 
An FDIA is perfectly undetectable by the controller if: 
\begin{equation}
    \bm{x}(t,\bm{x}(0),\bm{u},\bm{1},\bm{1})=\tilde{\bm{x}}(t,\bm{x}(0),\tilde{\bm{u}},\bm{\alpha},\bm{\beta})
    \label{perfectFDIAcontroldef}
\end{equation}
\noindent where $\tilde{\bm{x}}=\bm{\alpha}(\bm{x})$, $\tilde{\bm{u}}=\bm{\beta}(\bm{u})$, and $\bm{1}$ is the identify function (representing no attack). 

{\bf Remark 1 (Undetectable affine transformation-based FDIAs)}.
FDIAs illustrated in Fig. \ref{FDIA_manipulator} are defined to be perfectly undetectable if:
$\bm{q}(t,\bm{x}(0),\bm{u},\bm{1},\bm{1},\bm{1},\bm{1})=\tilde{\bm{q}}(t,\bm{x}(0),\bm{\tilde{u}},\bm{S}_x,\bm{d}_x, \bm{S}_u, \bm{d}_u)$
where $\bm{q}=\bm{1}(\bm{q})$, $\bm{u}=\bm{1}(\bm{u})$ are the identify functions.

It should be mentioned that simultaneous FDIA on the commands and observables is not necessarily a new concept. Papers on covert attacks in the literature introduced a similar structure in which the attacker implements an additional dynamic controller between the commands and observables as a stealthy attack \cite{Schellenberger}. In contrast, this paper formulates perfectly undetectable FDIAs in terms of affine transformations, as described in next section. 
Also, note that the prior work assumed $\bm{x}(0)=0$ for linearity assumptions. However, this paper derives the conditions including the one to match the observation of the initial condition for undetectable attacks. 
For the manipulator kinematic control, we do not assume $\bm{q}(0)=0$ since
$\tilde{\bm{q}}(0)=\bm{S}_x \bm{q}(0)+\bm{d}_x$
does not always hold.

\section{Realization of perfectly undetectable FDIA}

\subsection{Equivalent plant dynamics}

{\bf Remark 2 (Exposure of the controller information and the desired trajectory to the attacker)}.
Equation (\Ref{perfectFDIAcontroldef}) may be achieved by the attacker without the knowledge of controller $\bm{k}$. Indeed, typical attack detector (such as an state observer or an adaptive algorithm \cite{Zhang23}) attempts to detect an attack based on the observed dynamic relationship from $\bm{u}$ to $\bm{\tilde{x}}$, where the control scheme is commonly outside of the procedure. In other words, if an FDIA is implemented such that the observed dynamics of an attacked plant, the RHS of (\ref{perfectFDIAcontroldef}), is equivalent to the  dynamics of the nominal plant, the LHS of (\ref{perfectFDIAcontroldef}), the attack cannot be detected regardless of the feedback control scheme and the desired trajectory. While this cannot be easily extended to general control systems, certain dynamic plants, including the one shown in Fig. \ref{FDIA_manipulator}, enable such FDIAs.

\subsection{Manipulator kinematic control and its attackability}

Recall the manipulator nominal dynamics (i.e., $\tilde{\bm{u}}=\bm{u}$) with joint velocity control, 
\begin{equation}
   \dot{\bm{q}}=\bm{u}, \label{nominaldynamics}
\end{equation}
one would observe that, in many industrial motion control systems, the control of joint angles is fully decoupled across individual joints and exhibits linear (first-order) time-invariant dynamics. Control commands are represented by $\dot{\bm{q}}=\tilde{\bm{u}}=\bm{S}_u \bm{u}+\bm{d}_u$ and plant observables by $\tilde{\bm{q}} = \bm{S}_x \bm{q}+\bm{d}_x$. One can differentiate the latter, and plug in the former to achieve the resultant, observed attacked plant dynamics,
\begin{equation}
    \dot{\tilde{\bm{q}}} =\bm{S}_x \bm{S}_u \bm{u}+\bm{S}_x \bm{d}_u,
    \label{attackeddynamics}
\end{equation}
yielding the following theorem. 

{\bf Theorem 1: Perfectly undetectable FDIA on manipulator control system. }
The vector fields defined by (\ref{nominaldynamics}) and (\ref{attackeddynamics}) are equivalent and, thus, resultant observables are indistinguishable including the initial condition regardless of the control scheme if the following conditions are satisfied: 

\begin{itemize}
  \item Condition 1: $\bm{S}_x \bm{S}_u=\bm{I}^{n \times n}$
  \item Condition 2: $\bm{q}(0)=\bm{S}_x \bm{q}(0)+\bm{d}_x$
  \item Condition 3: $\bm{d}_u=0$.
\end{itemize}
\noindent {\it Proof}: Substituting Conditions 1 and 3 into (\ref{attackeddynamics}) yields (\ref{nominaldynamics}). Evaluating $\tilde{\bm{q}} = \bm{S}_x \bm{q}+\bm{d}_x$ at $t=0$ that must be identical to the nominal initial condition $\bm{q}(0)$ yields Condition 2.   $\blacksquare$

{\bf Remark 3 (Attack matrix selection)}. The attacker can utilize a range of $\bm{S}_x$ and $\bm{S}_u$ combinations such as scaling, reflection, shear, and rotation to satisfy Condition 1. In other words, this particular plant model is {\it highly attackable.}
The attacker needs to know the initial posture $\bm{q}(0)$ to satisfy Condition 2. Also, $\bm{d}_u=0$ is required as the nominal dynamics do not include an offset term. 

\subsection{Undetectable FDIA against adaptive attack detector} \label{subsec:zhang}

In \cite{Zhang23}, a specific case with $\bm{S}_u=\beta \bm{I}^{m \times m}, \bm{d}_u=0, \bm{S}_x= \bm{I}^{n \times n}, \bm{d}_x=0$ was considered with a scalar, constant attack parameter $0<\beta \leq 1$. Note that $\beta=1$ indicates no FDIA or normal operation.  The efficiency of the adaptive attack detector integrated into the Jacobian velocity control with projection operators, illustrated in Fig. \ref{FDIA_manipulator}, was demonstrated. The adaptive attack detector is represented as:
\begin{eqnarray}
    \dot{\hat{\bm{r}}}=\hat{\beta} \bm{J}(\tilde{\bm{q}}) \bm{u}-k_1 (\hat{\bm{r}}-f(\tilde{\bm{q}})) \label{dothatr}\\
    \dot{\hat{\beta}}=-k_2 \bm{u}^T \bm{J}(\tilde{\bm{q}})^T (\hat{\bm{r}}-f(\tilde{\bm{q}})) - s (\hat{\beta}-P_{\Gamma}(\hat{\beta})) \label{dothatbeta}
\end{eqnarray}
\noindent where $\hat{\beta}$ is the estimation of $\beta$. 
$P_{\Gamma}$ is a projection operator  \cite{proj} to maintain $\hat{\beta}$ within $\Gamma = \{\hat{\beta} \in R \ | \varepsilon \leq \hat{\beta} \leq 1\}, \varepsilon>0$ via a mapping function $s$. Similarly, $P_{\Omega}$ is another projection operator to maintain joint variables within specified displacement and velocity limits. 
$\hat{\bm{r}}$ monitors if the remote manipulator moves following the control command $\bm{u}$.
Note that (\ref{dothatr}) and (\ref{dothatbeta}) already include the compromised observable $\tilde{\bm{q}}$.
The proof of convergence of $\hat{\beta} \rightarrow \beta, t \rightarrow \infty$ was given by the Lyapunov function analysis  \cite{Zhang23}. 

In the FDIA discussed in this paper, the assumption that measurements of ground truth obvervables, or that $\bm{S}_x= \bm{I}^{n \times n}, \bm{d}_x=0$, no longer holds.
From Condition 1,  $\bm{S}_x=1/\beta \bm{I}^{n \times n}$ for $\bm{S}_u=\beta$. Assume $\bm{d}_x$ is chosen appropriately. 

{\bf Theorem 2: Undetectable FDIA by adaptive attack detector.} For the attack detector given (\ref{dothatr}) and (\ref{dothatbeta}), $\hat{\beta} \rightarrow 1, t \rightarrow \infty$, if $\bm{S}$ and $\bm{S}_u$ are chosen as $\bm{S}_x=1/\beta$ and $\bm{S}_u=\beta$. 

\noindent
{\it Proof}:
Let $\bm{r}'=f(\bm{\tilde{q}})$ be the end-effector position of the manipulator that the controller reconstructs from the compromised observables $\tilde{q}$.
If $\bm{r}'$ is identical to that expected from nominal control commands $\bm{u}$, the attack detector is unable to estimate $\beta$. Let $\tilde{\bm{r}}=\hat{\bm{r}}-f(\bm{\tilde{q}})$. From (\ref{dothatr}),
\begin{equation}
    \dot{\hat{\bm{r}}} = \hat{\beta} \bm{J}(\tilde{\bm{q}}) \bm{u}-k_1 \tilde{\bm{r}}
\end{equation}

\begin{eqnarray}
   \dot{\tilde{\bm{r}}}&=&\dot{\hat{\bm{r}}}-\frac{\partial f(\tilde{\bm{q}})}{\partial \tilde{\bm{q}}} \frac{d \tilde{\bm{q}}}{dt} 
= \dot{\hat{\bm{r}}} - \bm{J}(\tilde{\bm{q}}) \dot{\tilde{\bm{q}}} \nonumber\\
&=& \hat{\beta} \bm{J}(\tilde{\bm{q}}) \bm{u}-k_1 \tilde{\bm{r}} - \bm{J}(\tilde{\bm{q}}) \dot{\tilde{\bm{q}}}\nonumber\\
&=& \hat{\beta} \bm{J}(\tilde{\bm{q}}) \bm{u}-k_1 \tilde{\bm{r}} - \bm{J}(\tilde{\bm{q}}) \bm{u}\nonumber\\
&=& (\hat{\beta}-1) \bm{J}(\tilde{\bm{q}}) \bm{u}-k_1 \tilde{\bm{r}}  \label{dottilder}
\end{eqnarray}
\noindent
The last equation is derived as follows: $\beta \dot{\tilde{\bm{q}}}=\dot{\bm{q}}$ since $\beta \tilde{\bm{q}}=\bm{q}$. Also, $\dot{\bm{q}}=\tilde{\bm{u}}=\beta \bm{u}$. Therefore, $\dot{\tilde{\bm{q}}}=\dot{\bm{q}}/\beta=\beta/\beta \bm{u} = \bm{u}$. Note that (\ref{dottilder}) is a special case when $\beta=1$ of Equation (18) in \cite{Zhang23} ($\beta$ to be estimated has vanished in the adaptive control law), leading to $\hat{\beta} \rightarrow 1, t \rightarrow \infty$ regardless of $\beta$.  $\blacksquare$

Since $\dot{\bm{r}}'=\bm{J}(\bm{\tilde{q}})\dot{\tilde{\bm{q}}} = \bm{J}(\bm{\tilde{q}}) \bm{u}$, the controller perceives that $\bm{r}'$ is realized by following the nominal dynamics despite the actual trajectory $\bm{r} \neq \bm{r}'$, enabling a perfectly undetectable FDIA.  Note that (\ref{dottilder}) is derived for any $\bm{S}_x$ and $\bm{S}_u$ satisfying $\bm{S}_x \bm{S}_u=\bm{I}^{n \times n}$ (Theorem 1, Condition 1) without a loss of generality, indicating that residual-based attack detectors such as (\ref{dothatbeta}) are susceptible to perfectly undetectable FDIAs.

\section{Experiments}


The perfectly undetectable attack game to validate the presented results is executed on a 6 degree-of-freedom nonredundant maniplator (FANUC lR Mate 200iD/7L) through robot control software RoboDK with MATLAB API. The MATLAB program calculates control commands and RoboDK performs the physical robot TCP/IP connection and command execution.
First, a nominal, no attack trial is demonstrated. Secondly, the attack detector, (\ref{dothatr}) and (\ref{dothatbeta}) in \ref{subsec:zhang} is implemented on a detectable attack to compare the following undetectable scenarios against. The objective of the attacker is to perform the following scenarios:
\begin{itemize}
    \item Scenario 1: Scaling attack
    \item Scenario 2: Reflection attack
    \item Scenario 3: Shear attack,
\end{itemize}
all while remaining completely undetectable according to Conditions 1--3. For all trials and attacks, a smiley face is desired to be drawn by the manipulator's end effector, with $\bm{q}(0) = [0, -10, 10, 0, 0, 0]^T$ degrees.

{\bf Nominal Trial:}
    The nominal trial demonstrates the desired trajectory outcome when the system is not attacked. As such, $\beta = 1, \bm{S}_{x} = I^{6 \times 6}, \bm{S}_{u} = I^{6 \times 6}, \bm{d}_{x} = 0, \bm{d}_{u} = 0$ (no attack).
    
{\bf Detectable Attack:}
    The chosen attack for the detectable trial is a scaling attack to the control input, where $\beta = 0.25, \bm{S}_{x} = \bm{I}^{6 \times 6}, \bm{S}_{u} = 0.25 \bm{I}^{6 \times 6}, \bm{d}_{x} = [0, 30, -30, 0, 0, 0]^T, \bm{d}_{u} = 0$. Since this attack does not satisfy conditions 1--3, it is considered detectable.
    
{\bf Scenario 1:}
     The attacker in this scenario intends to implement a scaling attack on the manipulator. As such, the attacker chooses $\beta = 0.25, \bm{S}_{x} = 0.25 \bm{I}^{6 \times 6}, \bm{S}_{u} = 4 \bm{I}^{6 \times 6}, \bm{d}_{x} = [0, 30, -30, 0, 0, 0], \bm{d}_{u} = 0$. Conditions 1--3 are satisfied by these static variable choices.
    
{\bf Scenario 2:}
    The attacker intends to reflect the manipulator trajectory. As such, the attacker chooses $\beta = -1, \bm{S}_{x} = - \bm{I}^{6 \times 6}, \bm{S}_{u} = - \bm{I}^{6 \times 6}, \bm{d}_{x} = [0, -20, 20, 0, 0, 0], \bm{d}_{u} = 0$. Conditions 1--3 are satisfied by these static variable choices.  
    
{\bf Scenario 3:}
    Finally, the attacker intends to shear the manipulator trajectory. By shearing the trajectory, the attacker is manipulating joint values by varying quantities instead of all by the same quantity. Shearing is performed by choosing upper triangular matrices,\\ 
\[
\hspace{-1em}
\begin{array}{cc}
\begin{subarray}{c}
\bm{S}_{x} =
\begin{bmatrix}
1 & 1 & 0 & 0 & 0 & 0 \\
0 & 1 & 1 & 0 & 0 & 0 \\
0 & 0 & 1 & 1 & 0 & 0 \\
0 & 0 & 0 & 1 & 1 & 0 \\
0 & 0 & 0 & 0 & 1 & 1 \\
0 & 0 & 0 & 0 & 0 & 1 \\
\end{bmatrix}
\end{subarray},\!\!\!\!
&
\begin{subarray}{c}
\bm{S}_{u} =
\begin{bmatrix}
1 & -1 & 1 & -1 & 1 & -1 \\
0 & 1 & -1 & 1 & -1 & 1 \\
0 & 0 & 1 & -1 & 1 & -1 \\
0 & 0 & 0 & 1 & -1 & 1 \\
0 & 0 & 0 & 0 & 1 & -1 \\
0 & 0 & 0 & 0 & 0 & 1 \\
\end{bmatrix}
\end{subarray}
\end{array},
\]
  $ \bm{d}_{x} = [-20, 10, 0, 0, 0, 0], \bm{d}_{u} = 0$. Conditions 1--3 are once again satisfied by these static variable choices.

\begin{figure}
    \centering
    \begin{subfigure}{0.6\columnwidth}
        \centering
        \includegraphics[width=\linewidth, trim={0 0 0 .25cm}, clip]{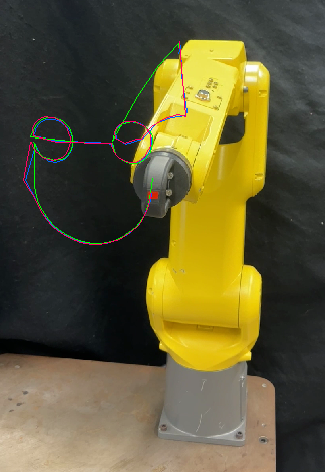}
        \caption{Nominal FANUC Results (See below for legend)}
        \label{subfig:nominal_a}
    \end{subfigure}
    \hfill
    \begin{subfigure}{.8\columnwidth} 
        \centering
        \includegraphics[width=\linewidth]{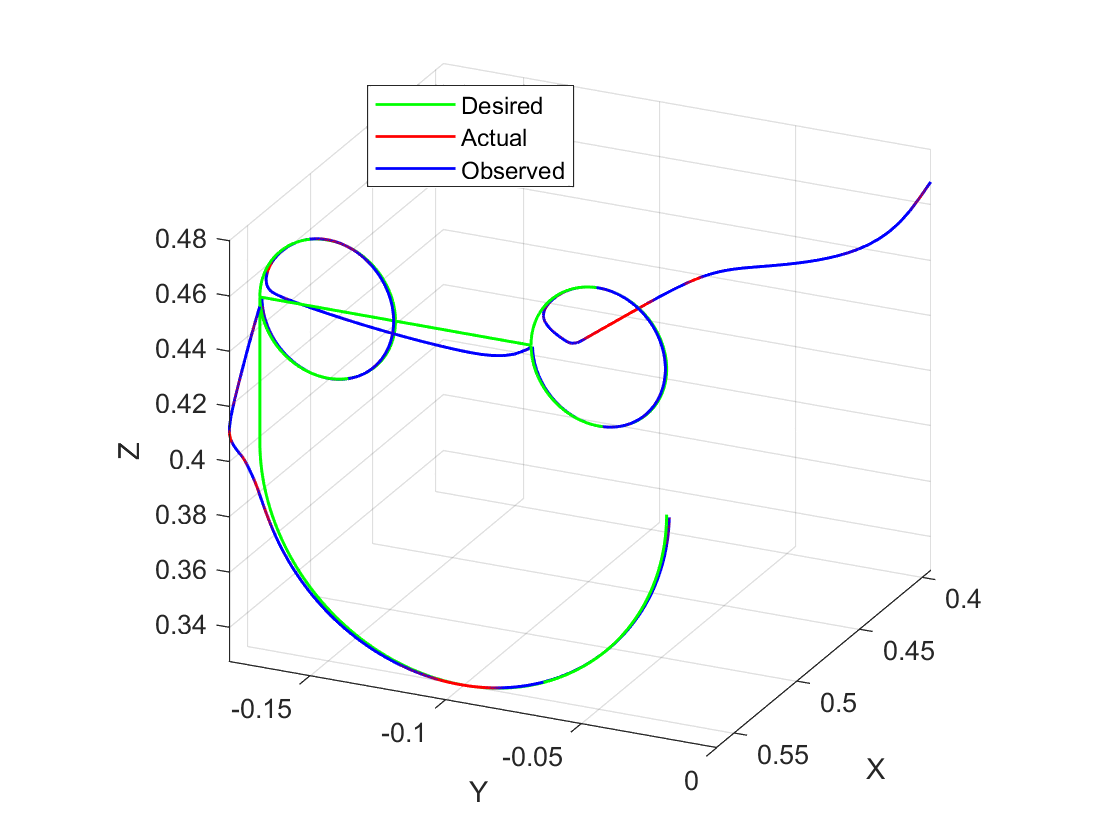}
        \caption{3D Trajectory (Meters)}
        \label{subfig:nominal_b}
    \end{subfigure}
    \\
    \begin{subfigure}{\columnwidth} 
        \centering
        \begin{subfigure}{.8\columnwidth}
            \centering
            \includegraphics[width=\linewidth]{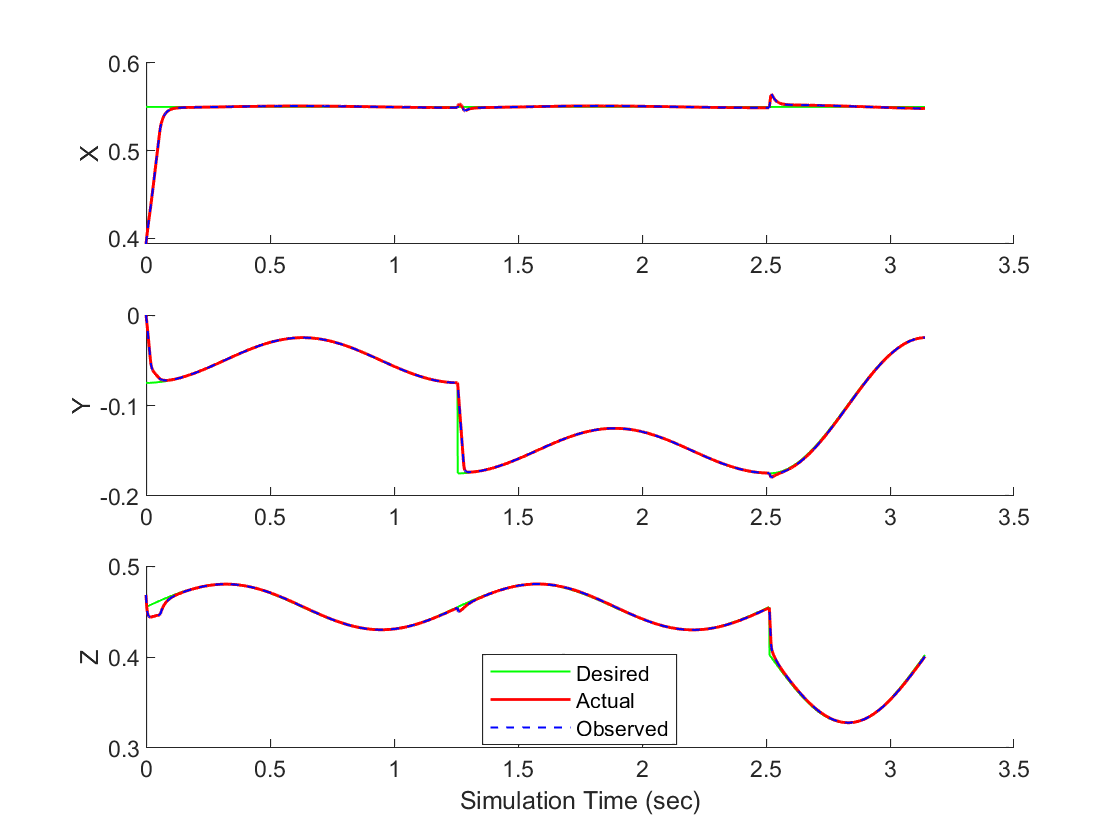}
            \caption{End Effector Position (Meters)}
            \label{subfig:nominal_c}
        \end{subfigure}
        \hfill
        \begin{subfigure}{.8\columnwidth}
            \centering
            \includegraphics[width=\linewidth]{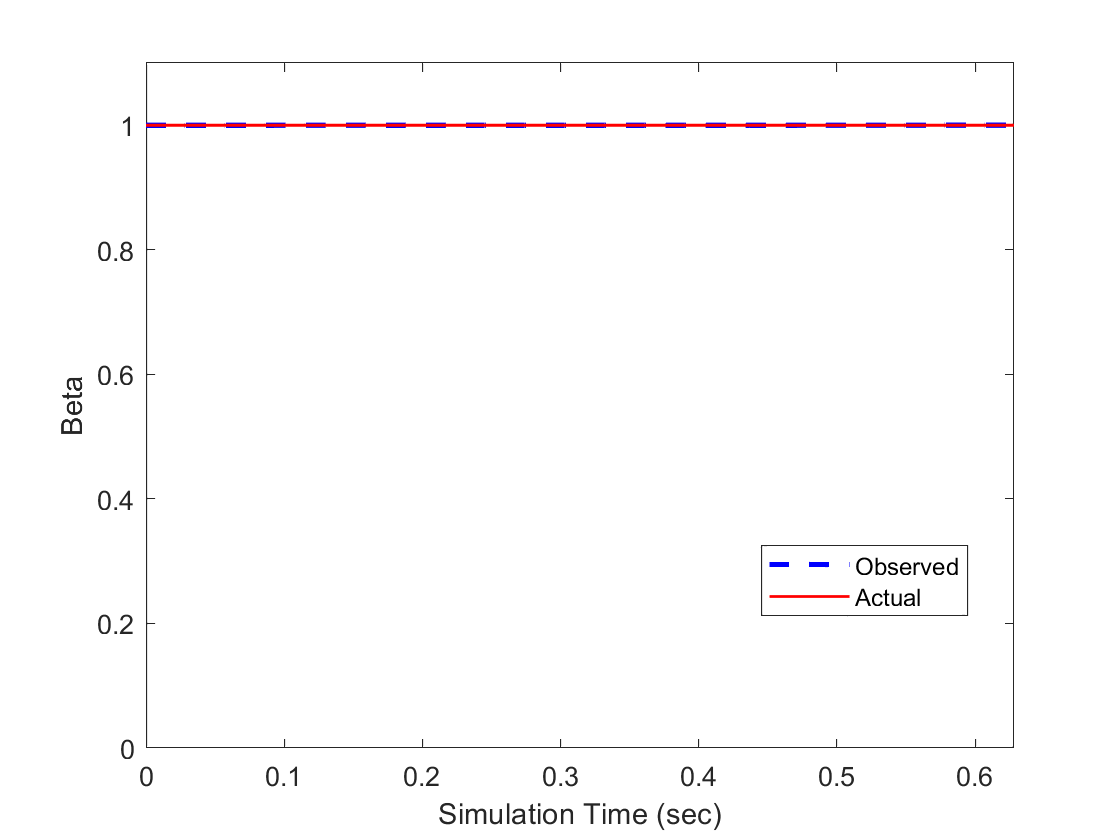}
            \caption{Beta Estimation}
            \label{subfig:nominal_d}
        \end{subfigure}
    \end{subfigure}
    \caption{Nominal Trial Results}
    \label{fig:nominal}
\end{figure}
\begin{figure}
    \centering
    \begin{subfigure}{0.6\columnwidth}
        \centering
        \includegraphics[width=\linewidth]{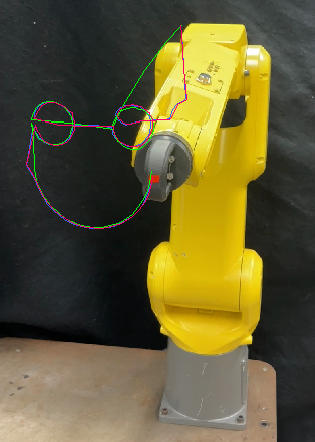}
        \caption{Detectable Attack FANUC Run (See below for legend)}
        \label{subfig:detectable_a}
    \end{subfigure}
    \hfill
    \begin{subfigure}{.8\columnwidth} 
        \centering
        \includegraphics[width=\linewidth]{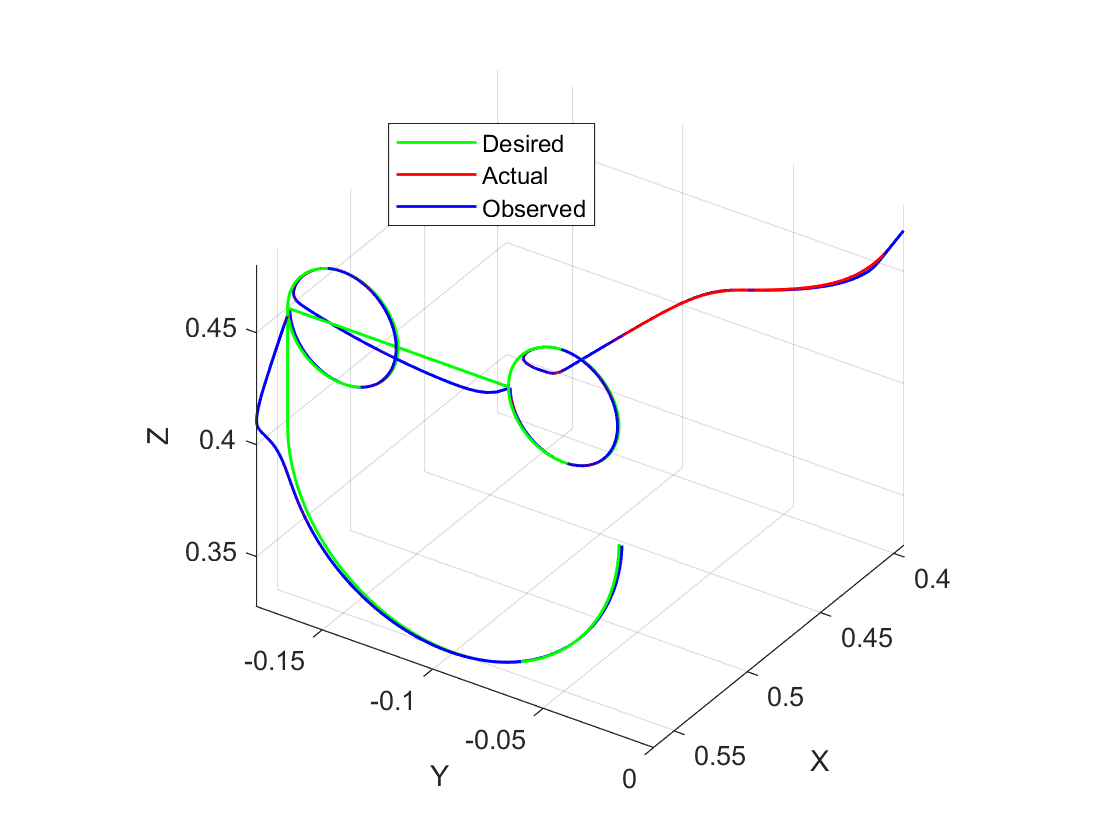}
        \caption{3D Trajectory (Meters)}
         \label{subfig:detectable_b}
    \end{subfigure}
    \\
    \begin{subfigure}{\columnwidth} 
        \centering
        \begin{subfigure}{.8\columnwidth}
            \centering
            \includegraphics[width=\linewidth]{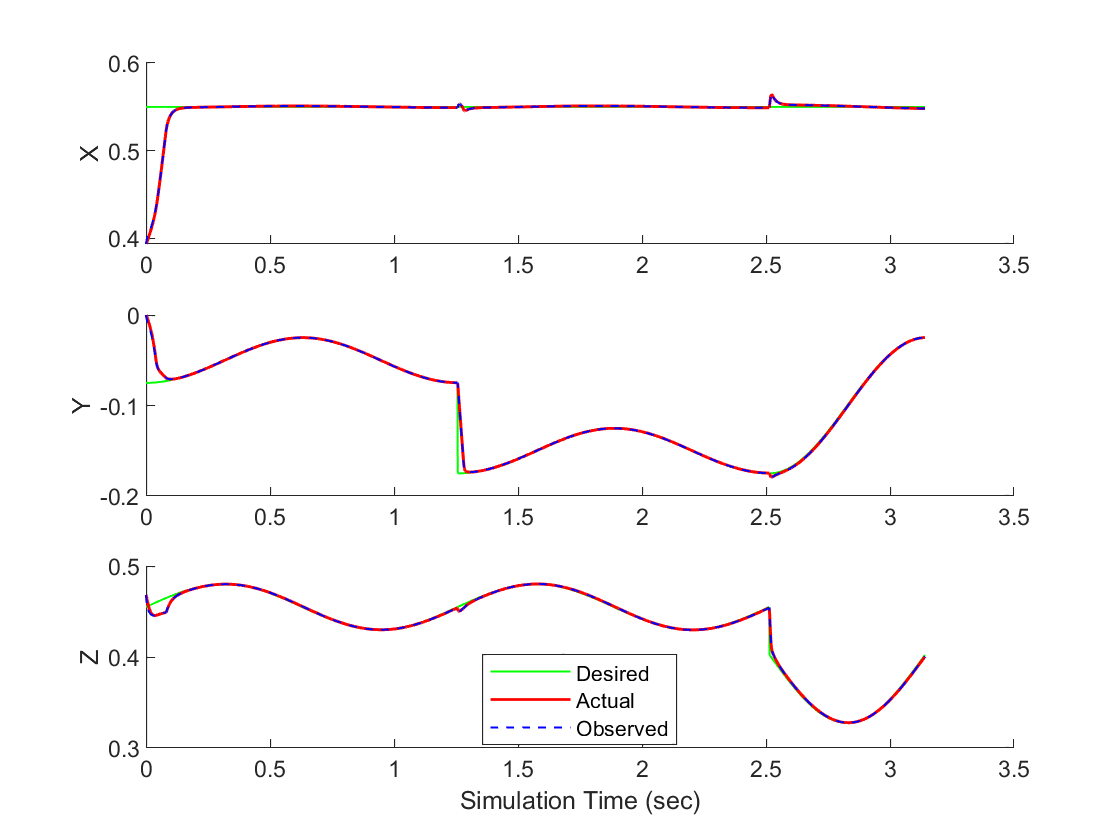}
            \caption{End Effector Position (Meters)}
             \label{subfig:detectable_c}
        \end{subfigure}
        \hfill
        \begin{subfigure}{.8\columnwidth}
            \centering
            \includegraphics[width=\linewidth]{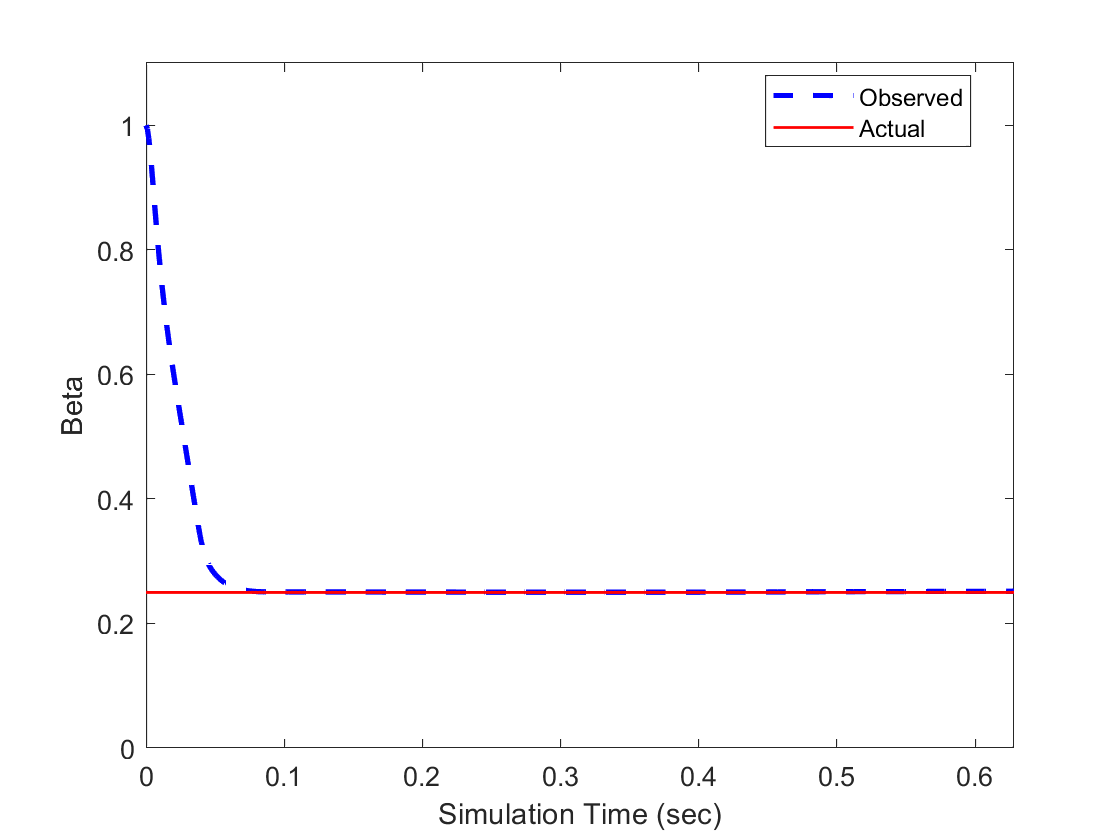}
            \caption{Beta Estimation}
            \label{subfig:detectable_d}
        \end{subfigure}
    \end{subfigure}
    \caption{Detectable Attack Results}
    \label{fig:detectable}
\end{figure}
\begin{figure}
    \centering
    \begin{subfigure}{0.6\columnwidth}
    \centering
    \includegraphics[width=\linewidth, trim={0 0.5cm 0 .25cm}, clip]{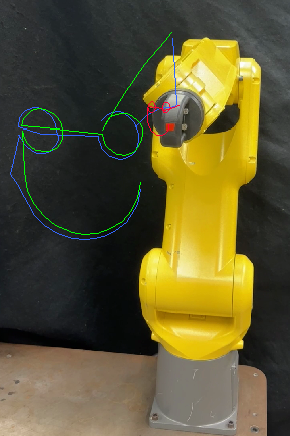}
    \caption{Scenario 1 FANUC Results (See below for legend)}
    \label{subfig:scale_a}
\end{subfigure}
    \hfill
    \begin{subfigure}{.8\columnwidth} 
        \centering
        \includegraphics[width=\linewidth]{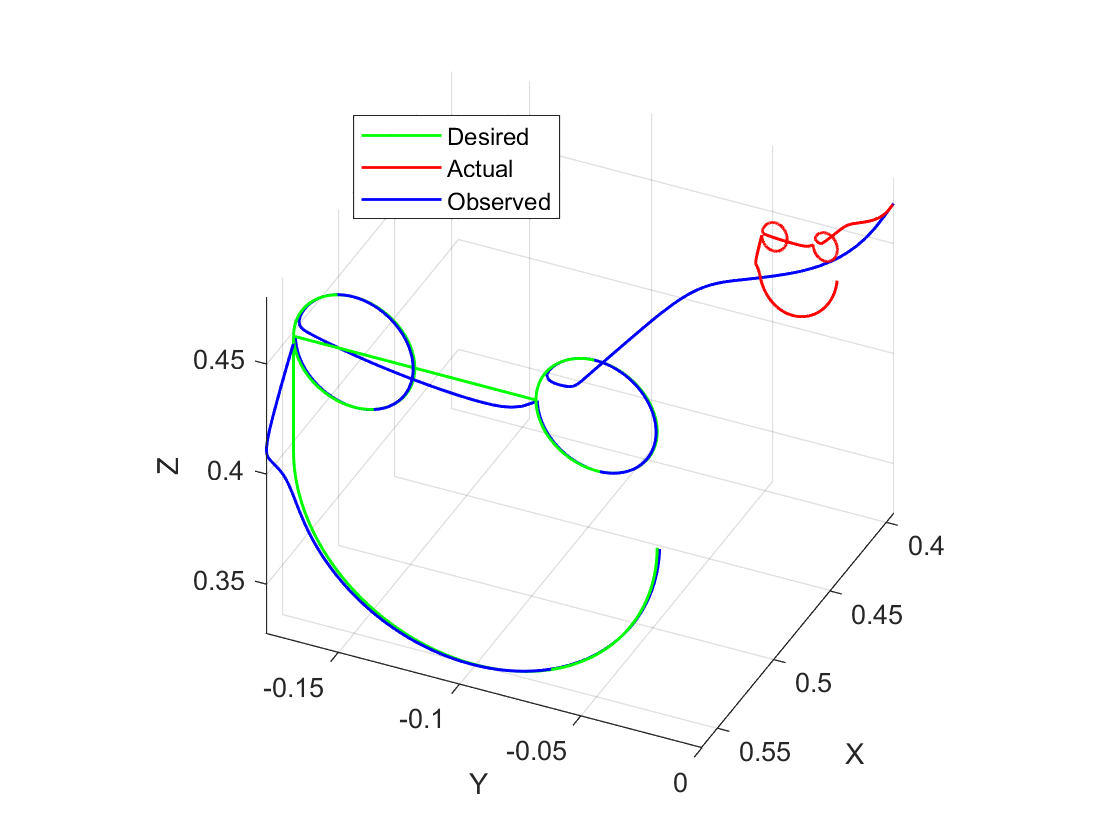}
        \caption{3D Trajectory (Meters)}
        \label{subfig:scale_b}
    \end{subfigure}
    \\
    \begin{subfigure}{\columnwidth} 
        \centering
        \begin{subfigure}{.8\columnwidth}
            \centering
            \includegraphics[width=\linewidth]{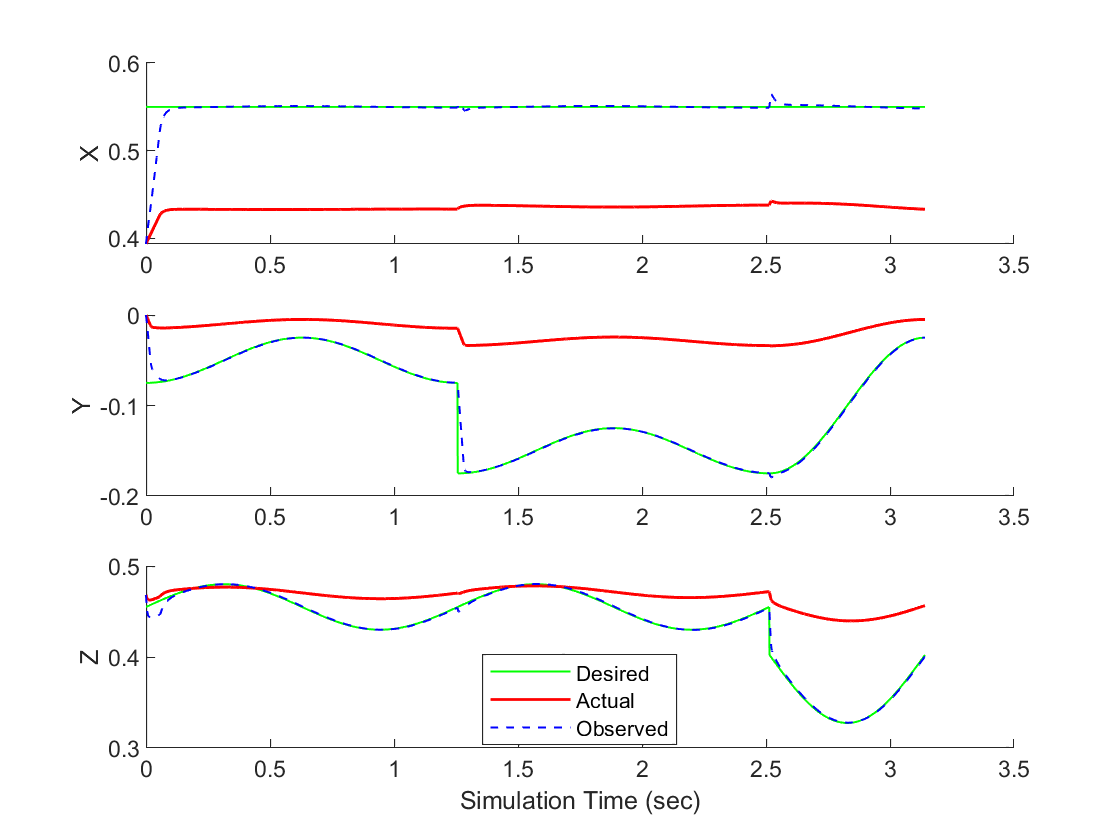}
            \caption{End Effector Position (Meters)}
            \label{subfig:scale_c}
        \end{subfigure}
        \hfill
        \begin{subfigure}{.8\columnwidth}
            \centering
            \includegraphics[width=\linewidth]{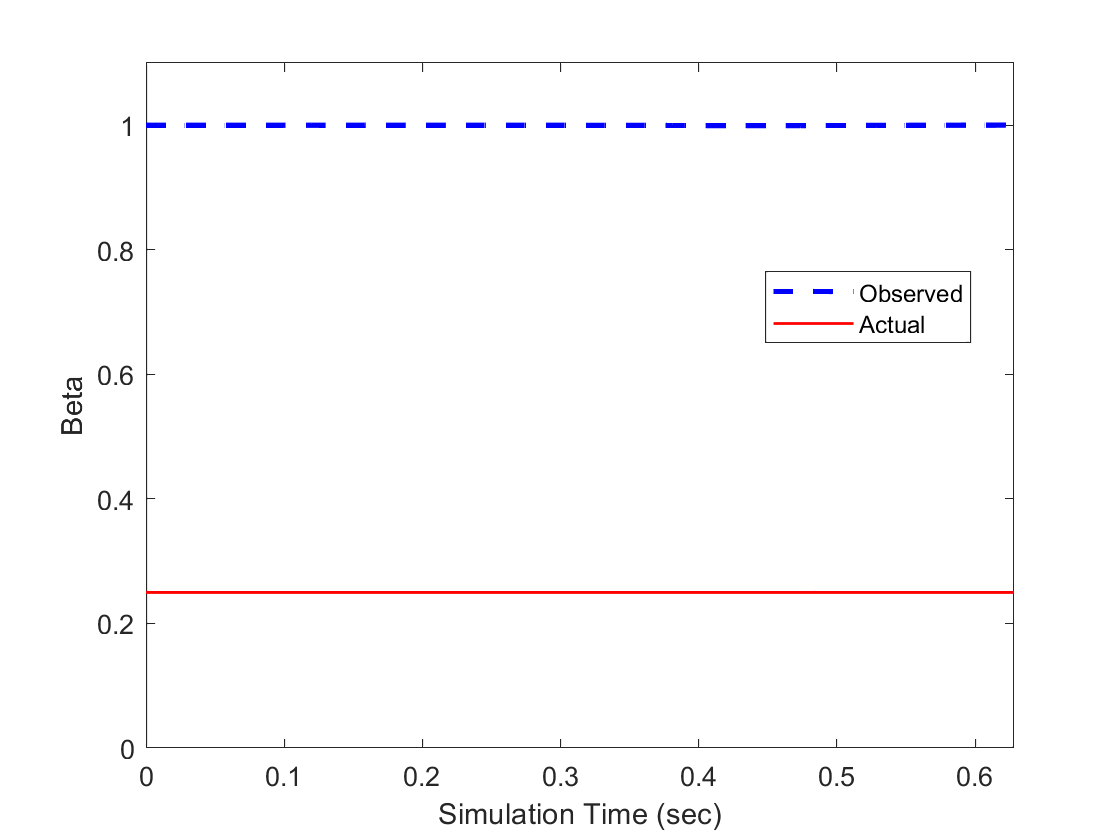}
            \caption{Beta Estimation}
            \label{subfig:scale_d}
        \end{subfigure}
    \end{subfigure}
    \caption{Scenario 1: Scaling Attack Results}
    \label{fig:scale}
\end{figure}
\begin{figure}
    \centering
    \begin{subfigure}{0.6\columnwidth}
        \centering
        \includegraphics[width=\linewidth, trim={0 .5cm 0 0}, clip]{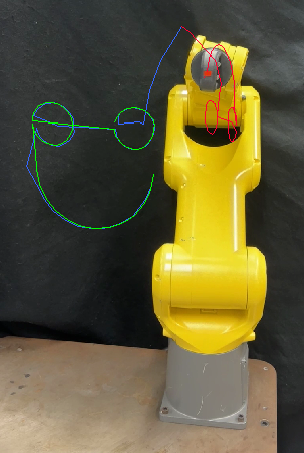}
        \caption{Scenario 2 FANUC Run (See below for legend)}
        \label{subfig:reflect_a}
    \end{subfigure}
    \hfill
    \begin{subfigure}{.8\columnwidth} 
        \centering
        \includegraphics[width=\linewidth]{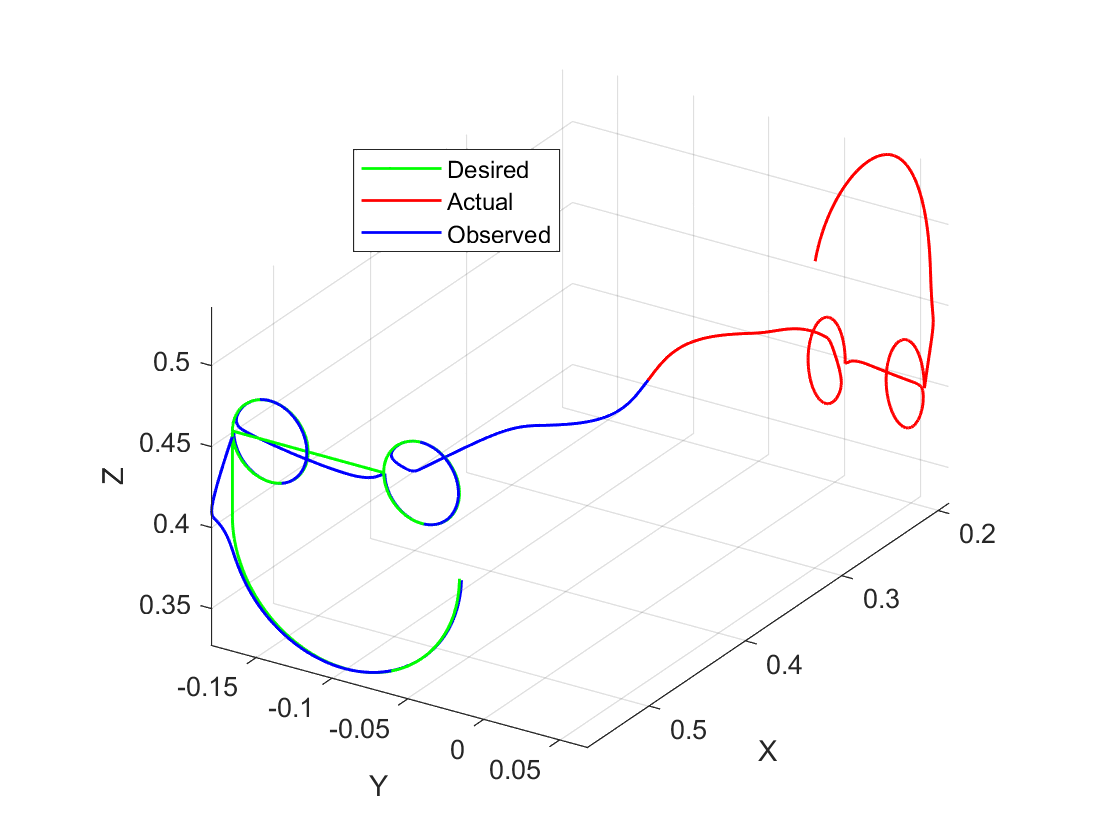}
        \caption{3D Trajectory (Meters)}
         \label{subfig:reflect_b}
    \end{subfigure}
    \\
    \begin{subfigure}{\columnwidth} 
        \centering
        \begin{subfigure}{.8\columnwidth}
            \centering
            \includegraphics[width=\linewidth]{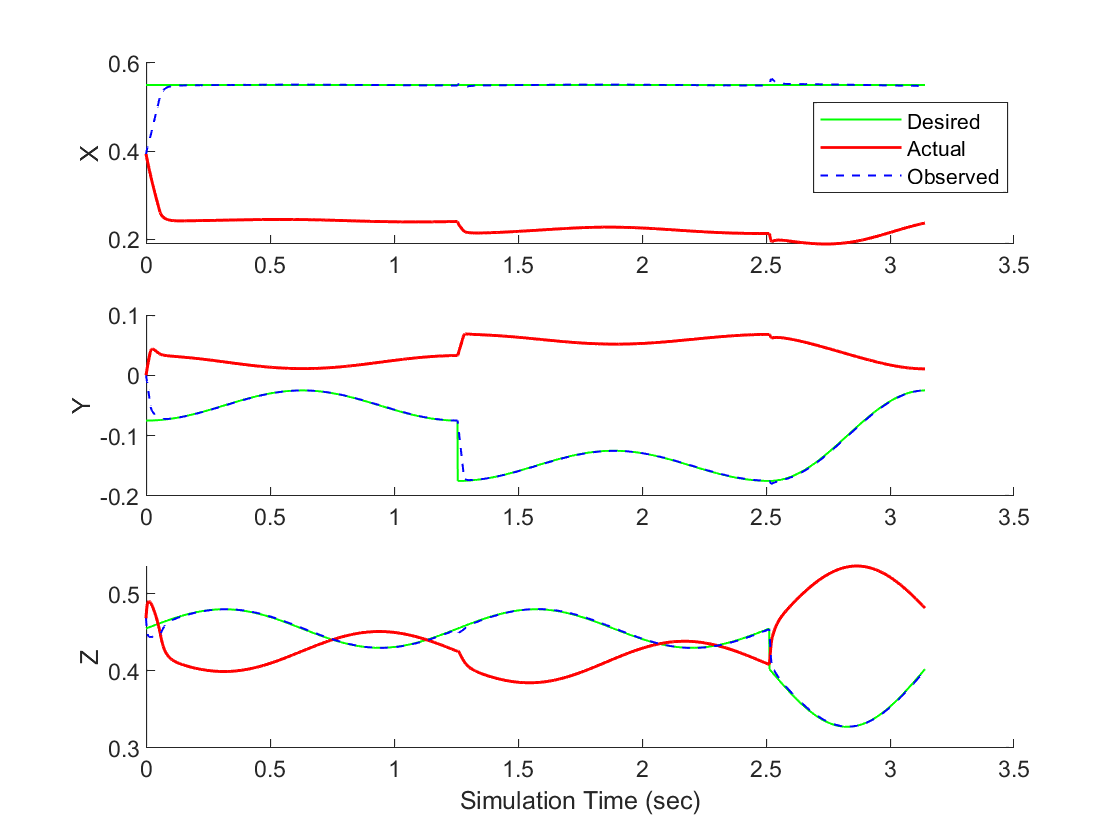}
            \caption{End Effector Position (Meters)}
             \label{subfig:reflect_c}
        \end{subfigure}
        \hfill
        \begin{subfigure}{.8\columnwidth}
            \centering
            \includegraphics[width=\linewidth]{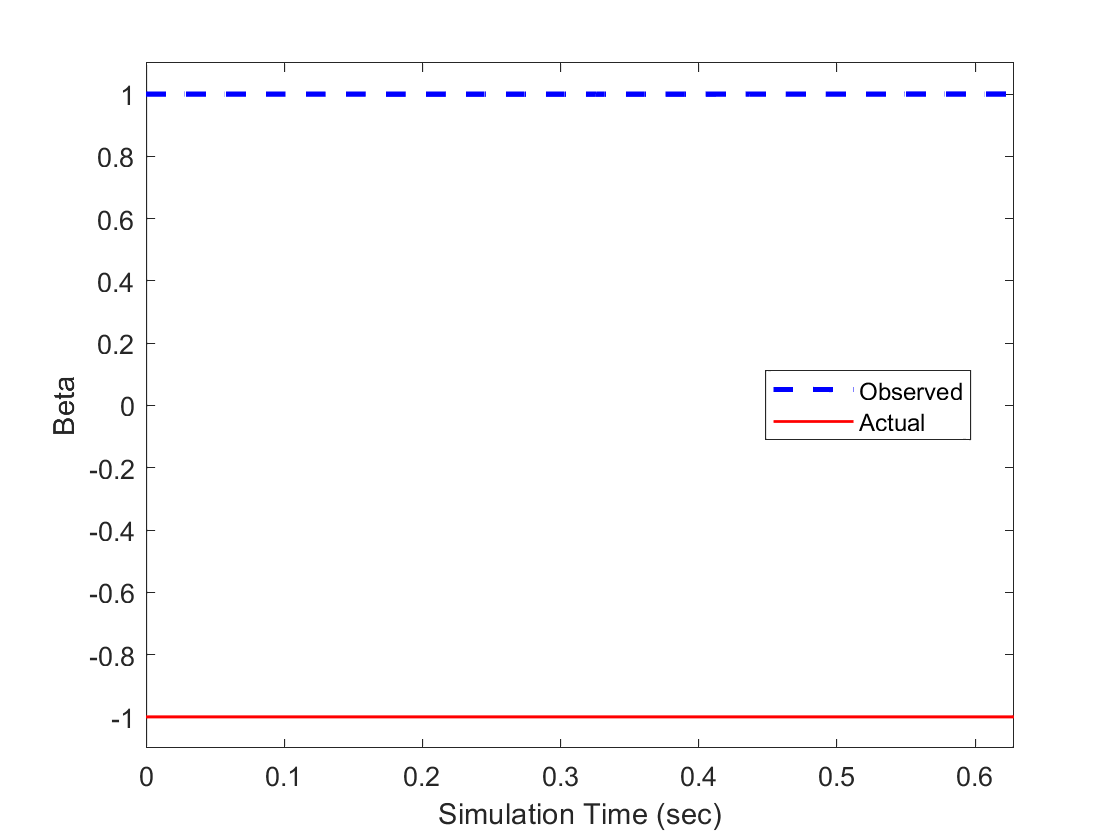}
            \caption{Beta Estimation}
             \label{subfig:reflect_d}
        \end{subfigure}
    \end{subfigure}
    \caption{Scenario 2: Reflection Attack Results}
    \label{fig:reflect}
\end{figure}
\begin{figure}
    \centering
    \begin{subfigure}{0.6\columnwidth}
        \centering
        \includegraphics[width=\linewidth, trim={0 1cm 0 0}, clip]{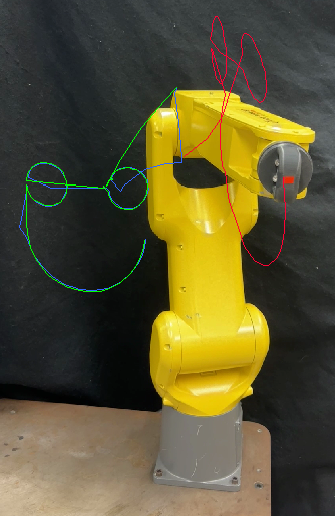}
        \caption{Scenario 3 FANUC Run (See below for legend)}
         \label{subfig:shear_a}
    \end{subfigure}
    \hfill
    \begin{subfigure}{.8\columnwidth} 
        \centering
        \includegraphics[width=\linewidth]{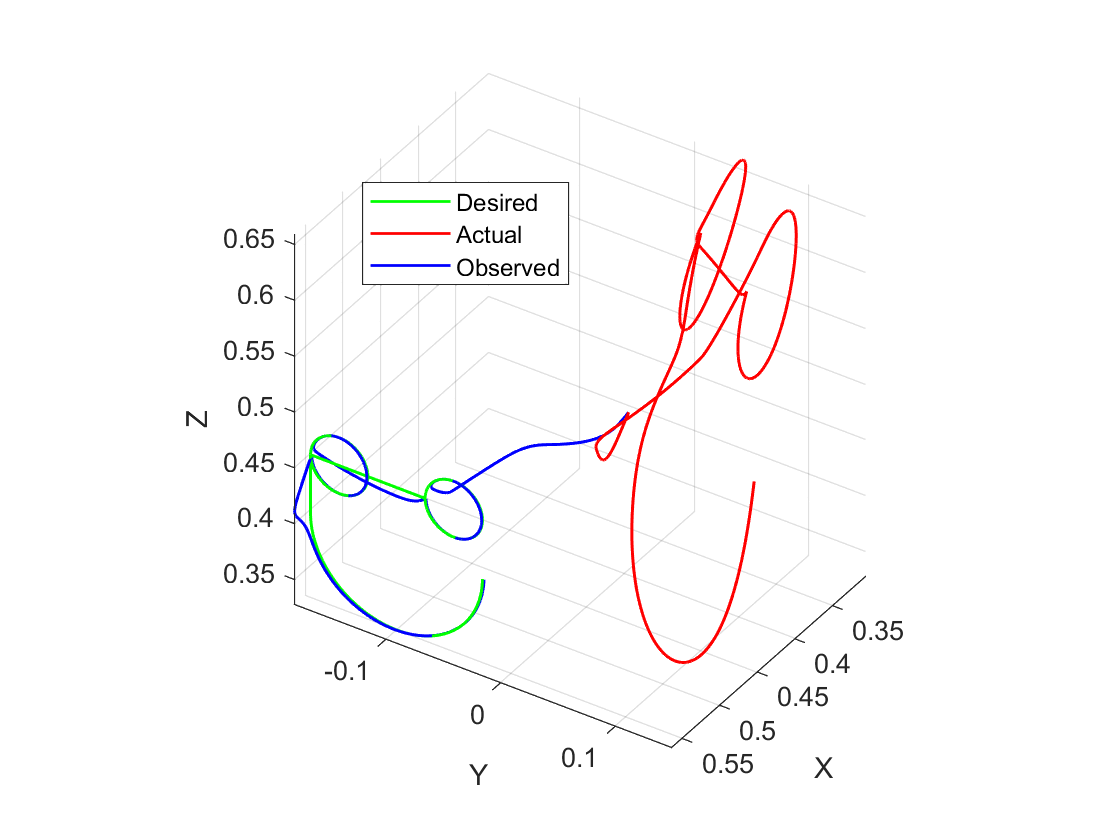}
        \caption{3D Trajectory (Meters)}
        \label{subfig:shear_b}
    \end{subfigure}
    \\
    \begin{subfigure}{\columnwidth} 
        \centering
        \begin{subfigure}{.8\columnwidth}
            \centering
            \includegraphics[width=\linewidth]{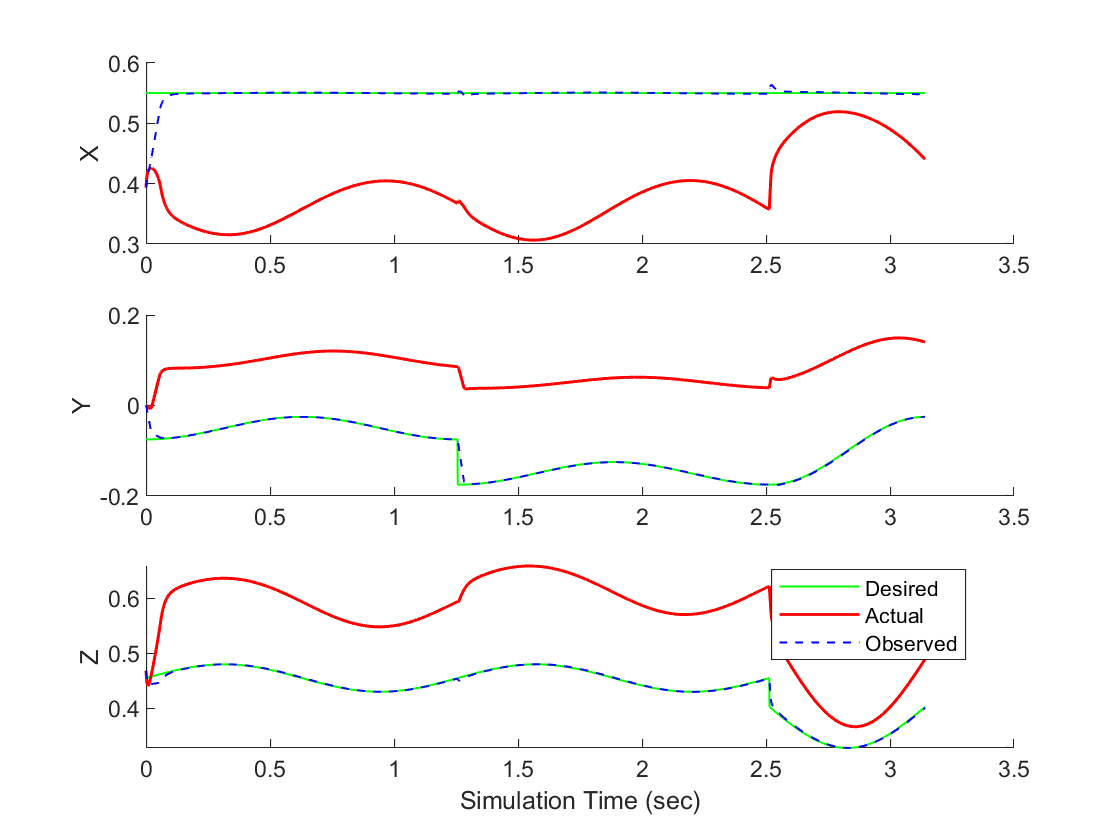}
            \caption{End Effector Position (Meters)}
            \label{subfig:shear_c}
        \end{subfigure}
        \hfill
        \begin{subfigure}{.8\columnwidth}
            \centering
            \includegraphics[width=\linewidth]{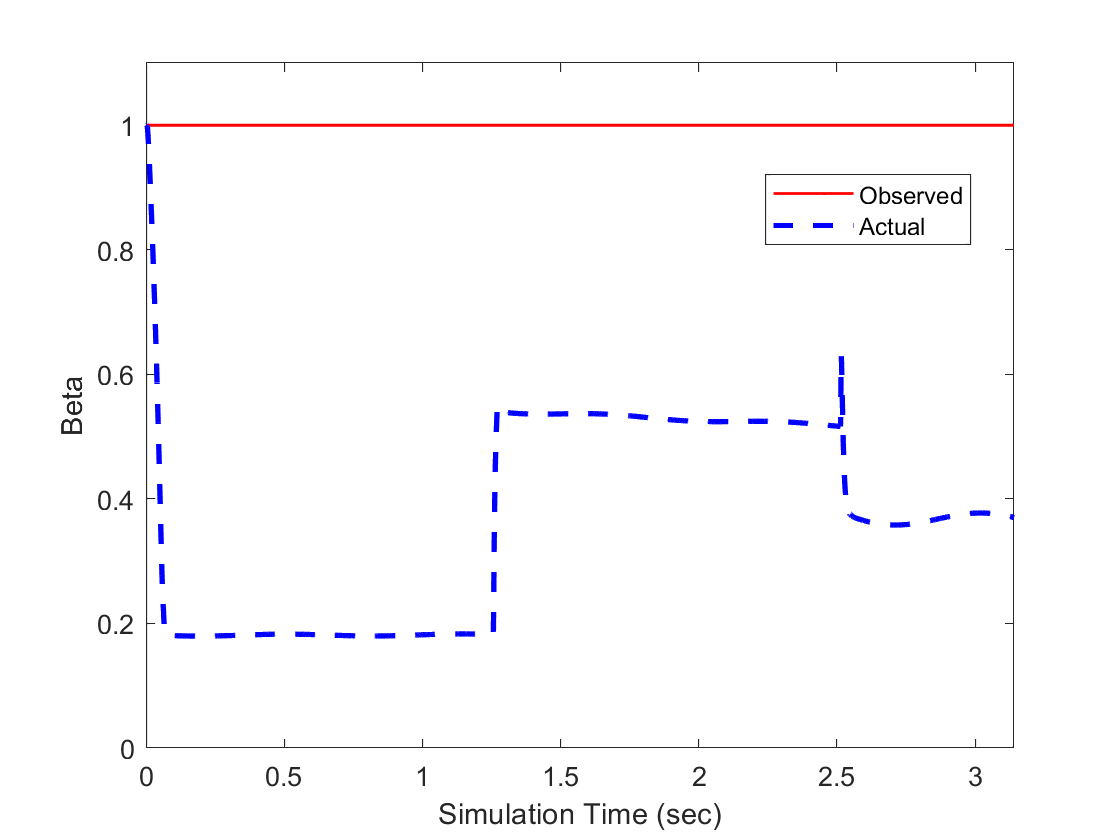}
            \caption{Beta Estimation - Note: Actual $\beta$ estimated by running detectable shear attack}
            \label{subfig:shear_d}
        \end{subfigure}
    \end{subfigure}
    \caption{Scenario 3: Shear Attack Results}
    \label{fig:shear}
\end{figure}

\begin{figure}[th]
	\centering
	\includegraphics[width=0.82\columnwidth]{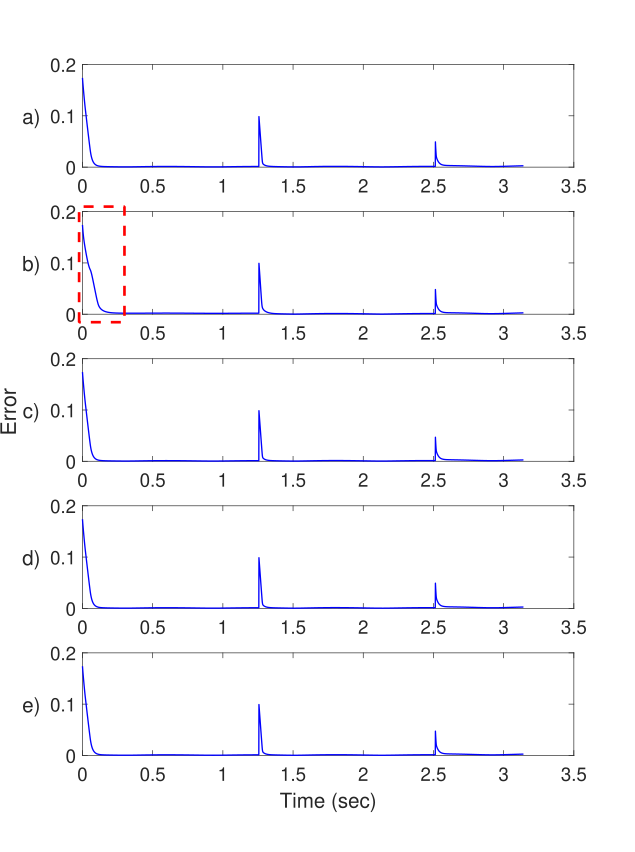}
	\caption{End Effector Observed Position Errors (Meters): a) Nominal Trial; b) Detectable Attack; c) Scenario 1: Scaling; d) Scenario 2: Reflection; e) Scenario 3: Shearing}
	\label{error}
\end{figure}

\section{Results and Discussions} \label{discussion}

Results are shown in Figs. \ref{fig:nominal}-\ref{fig:shear}. The nominal case demonstrates the smiley face drawn by the manipulator's end effector with no attack. As expected, the trajectory is accurately drawn via Jacobian transpose control with minimal error as demonstrated in Fig. \ref{error}a. The spikes in the error are the trajectories from one eye to the other and from the second eye to the mouth. For the detectable case in Fig. \ref{fig:detectable}, the attack detector converges the $\hat{\beta}$ to its actual value as in Fig. \ref{subfig:detectable_d}, and thus the manipulator draws the smiley face as desired even in the presence of an attack. The average error in the nominal case is $0.0017\,m$, but the detectable case is a bit larger at $0.0023\,m$, as can be seen in Fig. \ref{error}b's red, dashed box. This larger initial error is due to the transient period of $\hat{\beta}$'s convergence.

For Scenarios 1-3, upon satisfaction of Conditions 1-3, with knowledge of initial conditions, the attacker can perform undetectable attacks to manipulate kinematic manipulator trajectories as in Figs. \ref{fig:scale}-\ref{fig:shear}. According to the objective of the attacker, Scenarios 1-3 should have an identical observed error since the attacks are undetectable according to Remark 1, leaving the user observing seemingly perfect trajectory results. Figure \ref{error}c-d confirms this observation, as the observed errors perfectly agree with the nominal trial at $0.0017\,m$ average error. Although the observed errors are demonstrating proper operating procedure, the experimental results in Figs. \ref{subfig:scale_a}, \ref{subfig:reflect_a}, and \ref{subfig:shear_a} demonstrate the end effector drawing an unexpected smiley face. The attacker should ensure these manipulated trajectories  remain within the manipulator's operating workspace and do not move into singularity so that no faults are triggered on the operator's side. The controller does not have any effect on the outcome of the perfectly undetectable attack, and thus the attacker does not need any information about controller architecture. These virtual attacks represent physical modification of hardware; shearing attacks are a rerouting of sensing and actuator wiring such that joints operate as if they are a different joint. 

One of the primary requirements of the perfectly undetectable attack is that the plant, as seen by the controller, has a form of linear dynamics, i.e., individual joint velocity control. This is a common form of industrial robot control which is utilized in \cite{Zhang23} and in this experimentation. The presented manipulator dynamics is a simple linear integration in the joint space, which allows the effective attacks demonstrated in Scenarios 1-3. 
Oftentimes, a user may apply methods to compensate for nonlinearities in plant dynamics at a local plant, e.g., feedback linearization or compute torque method at the plant \cite{robotics} to simplify the remote controller. This architecture introduces a security hole by creating an easily attackable plant which may invite a variety of undetectable FDIAs. Note that the presence of nonlinearities in the plant does not fully prevent undetectable FDIAs; demonstration of perfectly undetectable attacks on nonlinear plants are currently being developed by the authors. 


\section{Conclusion}

This paper demonstrated that intelligent attackers can implement coordinated, perfectly undetectable FDIAs on control commands and observables in the form of affine transformations, allowing them to alter linear kinematic manipulator trajectories within their workspaces. Validation is provided through a nominal trial, a detectable case, and three Scenarios which show effective attacks on a  manipulator. Future work includes extension to nonlinear systems and development of an attack detection method as the defender. 

The authors would like to thank Dr. Kiminao Kogiso, the University of Electro Communications, for valuable discussions, and Dr. Yinyan Zhang, Jinan University College of Cyber Security, for their assistance on robot kinematic control and attack detection programming. This work was supposed in part by NSF CMMI Grant 2112793.

\appendices

 \section{Perfectly undetectable FDIA from the plant's perspective}
\label{appen_perfectFDIAplant}

{\bf Definition A1 (Perfectly undetectable FDIA from the plant's perspective)} (Milosevic 2021 \cite{Sandberg22,GRACY21}). Let $y(x(0),u,a)$ denote the response of the system for the initial condition $x(0)$, input $u(t)$, and attack signal $a(t)$. The attack is perfectly undetectable if 
\begin{equation}
    y(x(0),u,a)=y(x(0),u,0), t \geq 0.
    \label{perfectFDIAplantdef}
\end{equation}

The attacker does not leave any traces in the measurements of $y$, and can impact the system’s performance or behavior without being noticed by an attack detector that utilizes $y$ for attack detection. Research showed that (\ref{perfectFDIAplantdef}) can be achieved by zero dynamics attacks with the existence of transmission zeros \cite{Sandberg22,Milošević20,Mao20}. 
In this definition, the detector receives ground truth observables without being compromised, i.e.,  $\bm{S}_x=\bm{I}^{n \times n}, \bm{d}_x=\bm{0}$, as a special case of Fig. \ref{FDIA_manipulator}. 



\bibliographystyle{IEEEtran}
\bibliography{citations}
\end{document}